\documentclass[preprint,aps,nofootinbib,showpacs,12pt]{revtex4}
\usepackage{mathrsfs,amsmath,amssymb,amsfonts}
\usepackage{graphicx,graphics,epsfig,hhline,color}

\newcommand{\be}{\begin{eqnarray}}
\newcommand{\ee}{\end{eqnarray}}
\newcommand{\beq}{\begin{eqnarray}}
\newcommand{\eeq}{\end{eqnarray}}

\newcommand{\Z}{\mathbb{Z}}

\newcommand{\expmT}{e^{-\left(E_i-\mu\right)/T}}
\newcommand{\expmmT}{e^{-2\left(E_i-\mu\right)/T}}
\newcommand{\expmmmT}{e^{-3\left(E_i-\mu\right)/T}}
\newcommand{\exppT}{e^{-\left(E_i+\mu\right)/T}}
\newcommand{\exppppT}{e^{-3\left(E_i+\mu\right)/T}}
\newcommand{\expppT}{e^{-2\left(E_i+\mu\right)/T}}

\newcommand{\bc}{\begin{center}}
\newcommand{\ec}{\end{center}}

\begin{document}
\title{The QCD critical end point in the PNJL model}
\author{P. Costa}
\affiliation{Departamento de F\'{\i}sica, Universidade de Coimbra, 
P-3004-516 Coimbra, Portugal}
\author{C. A. de Sousa}
\affiliation{Departamento de F\'{\i}sica, Universidade de Coimbra, 
P-3004-516 Coimbra, Portugal}
\author{M. C. Ruivo}
\affiliation{Departamento de F\'{\i}sica, Universidade de Coimbra, 
P-3004-516 Coimbra, Portugal, EU}
\author{H. Hansen}
\affiliation{Univ.Lyon/UCBL, CNRS/IN2P3, IPNL ``Labo en lutte'' , 
4 rue E.Fermi, F-69622 Villeurbanne Cedex, France, EU}
\date{\today}

\begin{abstract}

We investigate the role played by the Polyakov loop in the dynamics of the chiral phase
transition in the framework of the so-called PNJL model in the $SU$(2) sector.
We present the phase diagram where the inclusion of the Polyakov loop moves the critical
points to higher temperatures, compared with the NJL model results.
The critical  properties of physical observables, such as the baryon
number susceptibility and the specific heat, are analyzed in the
vicinity of the critical end point, with special focus on their
critical exponents.
The results with the PNJL model are closer to lattice results  and we also recover
the universal behavior of the critical exponents of both the baryon susceptibility and
the specific heat.

\end{abstract}
\maketitle




Confinement and chiral symmetry breaking are two of the most important features of
quantum chromodynamics (QCD). Chiral models like the Nambu-Jona-Lasinio (NJL) model
have been successful in explaining the dynamics of spontaneous breaking of chiral
symmetry and its restoration at high temperatures and densities/chemical potentials.
Recently, this and other types of models, together with an intense experimental activity,
are underway to construct the phase diagram of QCD.

Results with two massless quarks in QCD show that, at high temperature, the phase
transition associated to restoration of chiral symmetry is second order and belongs to the
universality class of O(4) spin models in three dimensions \cite{Rajagopal:1993NPA}. With
small quark masses, the second order phase transition is replaced by a smooth crossover,
a picture which is consistent with lattice simulations \cite{Laermann:1998NPPS}.
Various results from QCD-inspired models indicate (see e.g. Refs.
\cite{Barducci:1994PRD,Stephanov:1996PRL}) that at low temperatures
the transition may be first order for large values of the chemical potential.
This suggests that the first order transition line may end when the temperature
increases, the phase diagram thus exhibiting a critical endpoint (CEP)
\cite{Schwarz:PRC1999,Buballa:2003PLB,Costa:2007PLB} that can be detected via
enhanced critical fluctuations in heavy-ion reactions
\cite{Stephanov:1998PRL,Hatta:2003PRD}.
At the CEP the transition is second order and belongs to
the Ising universality class \cite{Berges:1999NPB}.
In the chiral limit a tricritical point (TCP) is found in
the phase diagram, separating the second order transition line from the first-order one.

Recent developments in lattice QCD \cite{Karsch} indicate that the CEP is
likely to  be localized by a new  generation of experiments with relativistic nuclei
(CBM experiment at FAIR), suggesting to explore the  range of baryon number
chemical potential $\mu_B=100-500$ MeV.

In a previous work \cite{Costa:2007PLB},  in the framework of the Nambu-Jona-Lasinio
(NJL) model, we studied the phase diagram, focusing our attention on the CEP and the
physics near it, through the behavior of the baryon number susceptibility and the
specific heat.

In this work we study thermodynamic properties of strongly interacting matter using the
Polyakov-Nambu-Jona-Lasinio (PNJL) model. This extended model, first implemented  in
Ref. \cite{Meisinger:1996PLB}, provides a simple framework which couples the chiral
and the confinement order parameters.
The NJL model describes interactions between constituent quarks, giving the correct
chiral properties; static gluonic degrees of freedom are then introduced in the NJL
lagrangian through an effective gluon potential in terms of Polyakov loops
\cite{Meisinger:1996PLB,Ratti:2005PRD,Megias:2006PRD} with the aim of taking into account
features of both chiral symmetry breaking and deconfinement.  The  coupling of the quarks
to the Polyakov loop leads to the reduction of the weight of the quarks degrees of
freedom as the critical temperature is approached from above, which is interpreted as a
manifestation of confinement and is essential to reproduce lattice results.
We emphasize that the reduction of the weight of the quark degrees of freedom might
also have an important role for the critical behavior.

This effect should be more visible in the temperature domain, which can be explained by
the attractive interactions between the quarks and the effective gluon field which shifts
the chiral phase transition temperature to high values, allowing for a stronger first
order phase transition.

Hence it is demanding to use this improved NJL model to investigate relevant thermodynamical
quantities such as the CEP and the TCP.

Our  main goal is to locate the critical end point in the PNJL model \cite{Kashiwa:2007}
and confront the results with the NJL one and universality arguments.
Based on the fact that the CEP is a genuine thermodynamic singularity, being considered a
second order critical point, response functions like the specific heat and
susceptibilities can provide relevant signatures for phase transitions.
We notice that susceptibilities in general are related to fluctuations through the
fluctuation dissipation theorem, allowing to observe signals of phase transitions  in
heavy-ion reactions \cite{quat7,Nonaka:2005PRC}.

The Lagrangian of the $SU$(2)$\otimes$$SU$(2) quark model with
explicit chiral symmetry breaking where the quarks couple to a (spatially
constant) temporal background gauge field (represented in term of
Polyakov loops) is given by \cite{Pisa1,Ratti:2005PRD}:

\begin{eqnarray}
{\mathcal L_{PNJL}\,}&=& \bar q\,(\,i\, {\gamma}^{\mu}\,D_\mu\,-\,\hat m)\,q 
+ \frac{1}{2}\,g_S\,[\,{(\,\bar q\,q\,)}^2+\,\,{(\,\bar q \,i\,\gamma_5\,\vec{\tau}q\,)}^2\,]
- \mathcal{U}\left(\Phi[A],\bar\Phi[A];T\right).
\label{eq:lag}
\end{eqnarray}
The quark fields $q = (u,d)$ are defined in Dirac and color fields, respectively with two
flavors, $N_f=2$ and three colors, $N_c=3$, and  $\hat{m}=\mbox{diag}(m^0_u,m^0_d)$ is
the current quark mass
matrix. \\
The quarks are coupled to the gauge sector {\it via} the covariant
derivative $D^{\mu}=\partial^\mu-i A^\mu$. The strong coupling
constant $g_{Strong}$ has been absorbed in the definition of $A^\mu$:
$A^\mu(x) = g_{Strong} {\cal A}^\mu_a(x)\frac{\lambda_a}{2}$ where
${\cal A}^\mu_a$ is the $SU_c(3)$ gauge field and $\lambda_a$ are the
Gell--Mann matrices. Besides in the Polyakov gauge and at finite temperature
$A^\mu = \delta^{\mu}_{0}A^0 = - i \delta^{\mu}_{4}A^4$. \\
The Polyakov loop $\Phi$ (the order parameter of $\Z_3$ symmetric/broken phase transition
in pure gauge) is the trace of the Polyakov line defined by:
$ \Phi = \frac 1 {N_c} {\langle\langle \mathcal{P}\exp i\int_{0}^{\beta}d\tau\,
    A_4\left(\vec{x},\tau\right)\ \rangle\rangle}_\beta$.

The pure gauge sector is described by an effective potential
$\mathcal{U}\left(\Phi[A],\bar\Phi[A];T\right)$ chosen to reproduce at
the mean-field level the results obtained in lattice calculations:
\begin{eqnarray}
    \frac{\mathcal{U}\left(\Phi,\bar\Phi;T\right)}{T^4}
    =-\frac{b_2\left(T\right)}{2}\bar\Phi \Phi-
    \frac{b_3}{6}\left(\Phi^3+
    {\bar\Phi}^3\right)+ \frac{b_4}{4}\left(\bar\Phi \Phi\right)^2,
    \label{Ueff}
\end{eqnarray}
where
\begin{eqnarray}
    b_2\left(T\right)&=&a_0+a_1\left(\frac{T_0}{T}\right)+a_2\left(\frac{T_0}{T}
    \right)^2+a_3\left(\frac{T_0}{T}\right)^3 \nonumber
\end{eqnarray}
and $a_0 = 6.75,\,\, a_1 = -1.95,\,\, a_2 = 2.625,\,\, a_3 = -7.44,\,\, b_3 = 0.75,\,\, b_4 = 7.5,\,\,T_0=270$ MeV.

The parameters of the pure NJL sector are fixed at zero temperature as
in \cite{Buballa:2003PLB}: the three-momentum cutoff
used to regularize all the integrals is $\Lambda = 590$ MeV, $m^0_u m^0_d = 6$ MeV and $g_S\Lambda^2 = 2.435$. They yield $M_{vac} = 400$
MeV, $m_{\pi} = 140.2$ MeV, $f_{\pi} = 92.6$ MeV and
$\langle{\bar u}u\rangle^{1/3} = (-241.5 \mbox{MeV})^3$ .

Finally with $E^2_p=p^2+M^2$ the $SU(N_f=2)$ PNJL grand potential is given by
\cite{Ratti:2005PRD,Hansen:2007PRD}:
\begin{widetext}
\be
\Omega(\Phi, \bar\Phi, M ; T, \mu)
&=&{\cal U}\left(\Phi,\bar{\Phi},T\right)
+2g_{_{S}} N_f\left\langle\bar{q_{i}}q_{i}\right\rangle^2
- 2 N_c\,N_f \int_\Lambda\frac{\mathrm{d}^3p}{\left(2\pi\right)^3}\,{E_p}\nonumber \\
&-& 2N_f\,T\int_\Lambda\frac{\mathrm{d}^3p}{\left(2\pi\right)^3}
\bigg\{
   \ln\left[ 1 + 3\bar\Phi\expmT + 3\Phi\expmmT + \expmmmT \right] \nonumber\\
&+& \ln\left[ 1 + 3\Phi\exppT + 3\bar\Phi\expppT + \exppppT \right]
\bigg\}.
\label{omega}
\ee
\end{widetext}
We notice explicitly that at $T=0$ the Polyakov loop and the quark sector decouples.

The baryon number susceptibility and the specific heat are  the response of the
baryon number density $\rho_q(T,\mu)$ and the entropy $S (T,\mu)$ to an infinitesimal
variation of the quark chemical potential $\mu$ and temperature, given respectively by:
\begin{equation}
    \chi_q = \left(\frac{\partial \rho_q}{\partial\mu}\right)_{T}, \hskip1cm  {\rm and}
    \hskip1cm  C = \frac{T}{V}\left ( \frac{\partial S}{\partial T}\right)_{\mu}.
    \label{chi}
\end{equation}

The baryon number density is given by
$\rho_q = \frac{ N_c}{\pi^2}\int p^2 dp \left( n_q(\mu ,T) -\bar{n}_q(\mu ,T)\right)$
where $n_q(\mu ,T)$ and $\bar{n}_q(\mu ,T)$ are the occupation numbers modified
by the Polyakov loop \cite{Hansen:2007PRD}.


The PNJL thermodynamic potential is an effective potential depending on
three parameters: $M$, $\Phi$ and $\bar\Phi$.
These parameters are not independent (nor the corresponding phase transitions)
since they should verify the mean-field equations $\partial \Omega/\partial M = 0$ and
$\partial \Omega/\partial \Phi = \partial \Omega/\partial \bar\Phi = 0$. With the two
last equations one can compute $\Phi$ and $\bar\Phi$ as functions of $M$ for any value of
$T$ and $\mu$. Hence we consider that the thermodynamic potential is an effective potential
depending only on $M$.

A common feature shared by NJL and PNJL models is that
the thermodynamic potential may have two degenerate minima at which two
phases have equal pressure and chemical potential and can coexist according to the
Gibbs criterium. In fact, this pattern is characteristic of a first order phase
transition: the two minima correspond, respectively, to the phases of broken and restored
symmetry.
The quark condensate can be identified with the order parameter whose values
allow to distinguish the two coexisting phases.
As the temperature increases, the first order transition persists up to the CEP. At the
CEP the chiral transition becomes of second order.  For temperatures above the CEP the
thermodynamic potential has only one minimum and the transition is washed out: a smooth
crossover takes place.

So, while from a qualitative point of view the results of both models are similar, they
differ quantitatively in two aspects: in the PNJL, the CEP and TCP are pushed to higher
values of the temperature  and the size of the critical region is larger. Let us now
analyze those results in more detail.

\begin{figure*}[t]
\begin{center}
  \begin{tabular}{cc}
    \hspace*{-0.5cm}\epsfig{file=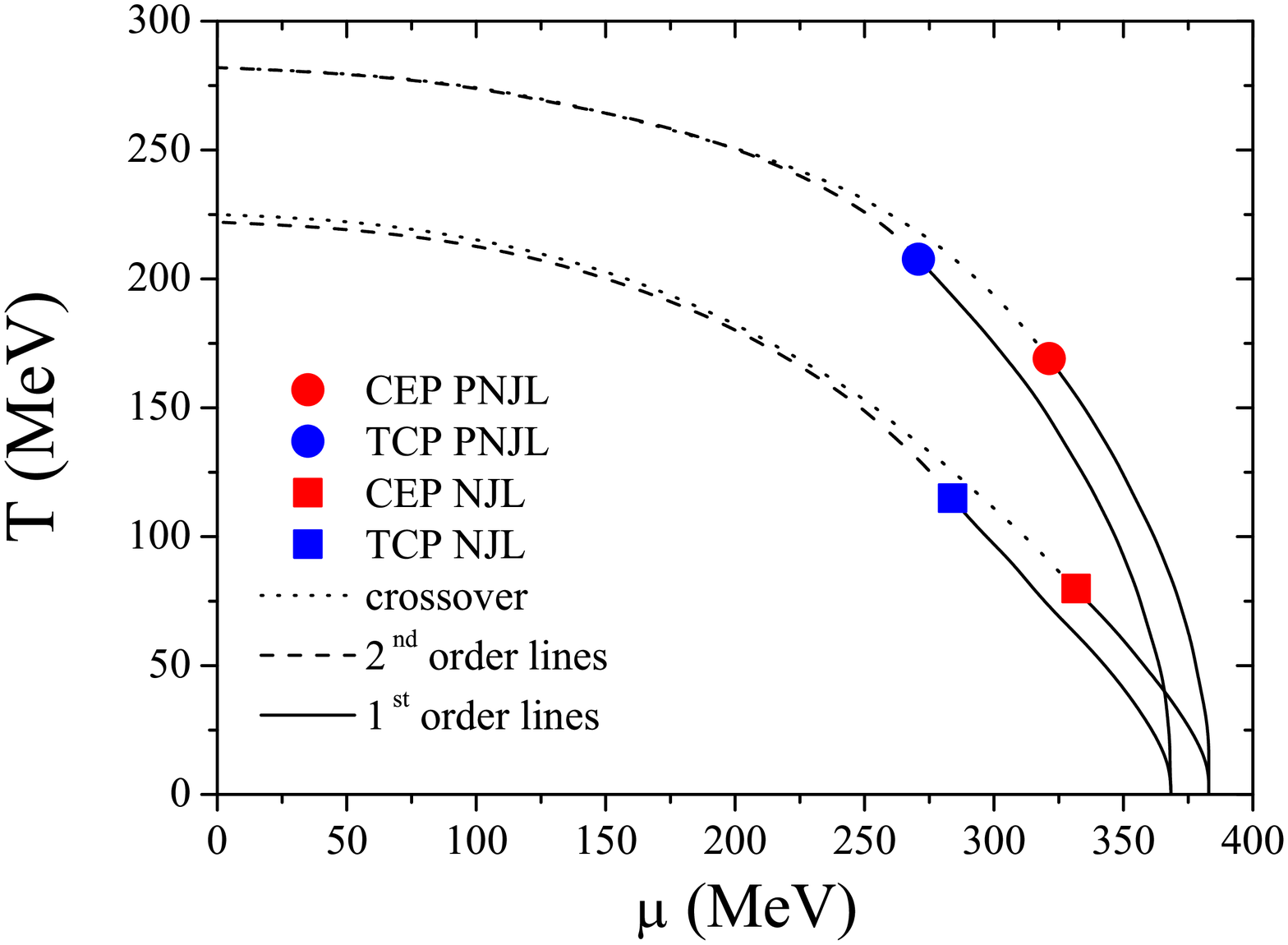,width=8.5cm,height=7.5cm} &
    \hspace*{-0.5cm}\epsfig{file=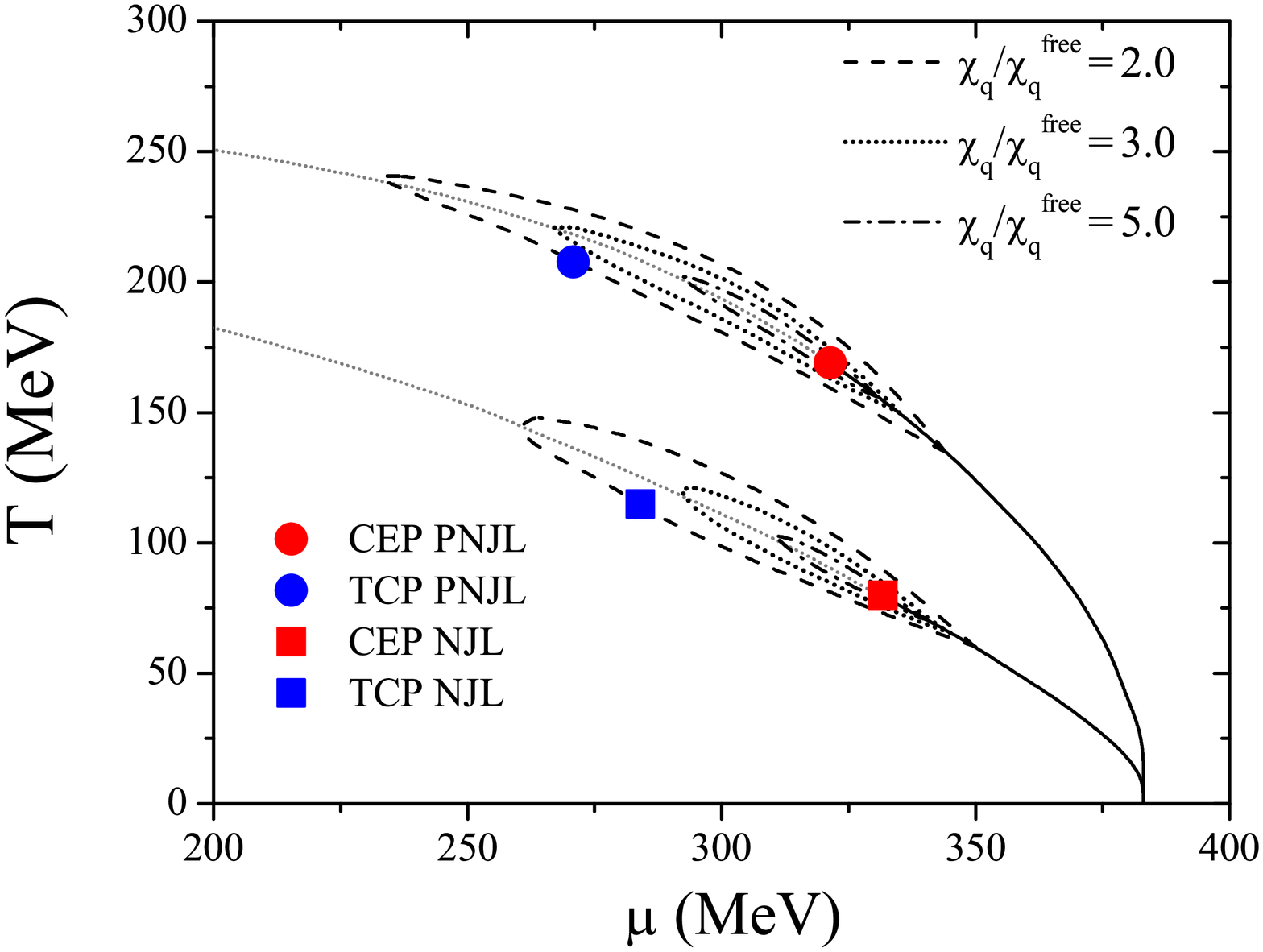,width=8.5cm,height=7.5cm} \\
   \end{tabular}
\end{center}
\vspace{-0.5cm} \caption{Left panel: the phase diagram in the PNJL and NJL models.
Right panel: the size of
the critical region is plotted for $\chi_q/\chi_q^{free}=2,3,5$.}
\label{Fig:1}
\end{figure*}

In the left panel of Fig. \ref{Fig:1} we plot the phase diagram for both,
PNJL and NJL models.
In the PNJL (NJL) model the CEP is localized at $T^{CEP}=169.11$ $(79.92)$ MeV and
$\mu^{CEP} = 321.32$ $(331.72)$ MeV ($\rho_q^{CEP}=2.76 (2.95) \rho_0$).
In the PNJL (NJL) model the TCP is located at
$T^{TCP}=207.66$ $(112.08)$ MeV and  $\mu^{TCP}=270.80$ $(286.05)$ MeV.
We remark that our  values for the CEP in the PNJL model are closer to the lattice results
of \cite{Karsch} and  the main change, with regards  to NJL values, is in $T^{CEP}$ and
$T^{TCP}$, a result which seems natural since the effects of the inclusion of the
Polyakov loop are expected to be more relevant in the domain of the temperature.
%

The size of the critical region around the CEP can be found by calculating the baryon
number susceptibility, the specific heat and their critical behaviors. The size of this
critical region is important for future searches of the CEP in heavy ion-collisions
\cite{{Nonaka:2005PRC}}.
To estimate the critical region around the CEP we calculate the dimensionless ratio
$\chi_q/\chi_q^{free}$, where $\chi_q^{free}$ is obtained taking the chiral limit $m^0 = 0$.
The right panel of Fig. \ref{Fig:1} shows a contour plot for three fixed ratios
($\chi_q/\chi_q^{free}=2.0,\,3.0,\,5.0$) in the phase diagram around the CEP
where we notice an elongation of the region where $\chi_q$ is enhanced,
in the direction parallel to the first-order transition line.
We also observe that the critical region is heavily stretched in the direction of the
crossover transition line as shown in Fig. \ref{Fig:1}.

The elongation of the critical region in the ($T$, $\mu$)-plane, along the critical line,
is larger in the PNJL model
(see $\chi_q/\chi_q^{free} = 2.0$ in the right panel of Fig. \ref{Fig:1}).
It means that the divergence of the correlation length at the CEP affects the phase
diagram quite far from the CEP and that a careful analysis including effects beyond the
mean field needs to be done \cite{Rossner:2007ik}.

As seen in \cite{Hansen:2007PRD} (Fig. 7), one of the main effects of the Polyakov loop is
to shorten the temperature range where the crossover occurs (at $\mu = 0$ the crossover
occurs within a range of 150 MeV  for the NJL model and within 115  MeV  for the PNJL
one), thus resulting in higher baryonic susceptibilities even far from the CEP.
This effect is driven by the fact that the one- and two-quark Boltzmann factors are
controlled by a factor proportional to $\Phi$: at small temperature, $\Phi \simeq 0$,
results in a suppression of these contributions. The thermal bath being then only
produced via the 3-quark Boltzmann factor, our physical interpretation is that the bath
is colorless, quarks being produced only in triplet necessarily colorless in the average
because of $\Phi$ being the order parameter of $\Z_3$ in this effective theory, $\Phi \simeq
0$ indicates a partial restoration of the color symmetry.  When the temperature
increases, $\Phi$ goes quickly to 1, resulting in a (partial) restoration of the chiral
symmetry which occurs in a shorter temperature range. In fact, the most striking
difference between NJL and PNJL models is a faster variation with temperature, around any
characteristic critical temperature, of the PNJL results.

The crossover taking place in a smaller temperature range can be interpreted
as a crossover transition closer to a second order one than in the NJL
model.
This ``faster'' crossover may also explain the elongation of the critical region
compared to the NJL one giving raise to a greater correlation length even far from
the CEP.

With this indication of the important role of the entanglement of the chiral and the
Polyakov loop dynamics on  the critical behavior of the QCD phase diagram, it is
mandatory to   investigate the behavior of $\chi_q$ and $C$ in the vicinity of the CEP
and their critical exponents, in the framework of the PNJL model. For comparison
purposes with the NJL model and the universality/mean-field predictions, the calculated
critical exponents at CEP and the TCP are presented in Table I, and will be discussed in
the sequel.

\begin{table}\label{Tab:1}
\begin {center}
\begin{tabular}{cccccccc}
    \hline
    {Quantity $\mbox{ }$  } & {  C.E./path} && { PNJL} && { NJL} && {Universality} \\
    \hline \hline
  & {$\epsilon\,/\,\,\rightarrow$\textcolor{red}{$\bullet$}} && { {$0.66 \pm 0.01$}}
  && {$0.66 \pm 0.01$} && {$2/3$} \\
  {$\chi_q$} & {{$\epsilon^\prime$\,/\,\,\textcolor{red}{$\bullet$}$\leftarrow$}} &&
  {$0.69 \pm 0.02$} && {$0.66 \pm 0.01$} && {$2/3$} \\
  &  $\gamma_q\,/\rightarrow$\textcolor{blue}{$\bullet$} &&
  $0.51 \pm 0.01$ && $0.51 \pm 0.01$ &&  {$1/2$}  \\
    \hline
    & {$\alpha\,/
        \begin{array}{c}
          \textcolor{red}{\bullet} \\
          \uparrow
        \end{array}$}
    && {$
        \begin{array}{c}
          \alpha=0.63\pm 0.02 \\
          \alpha_1=0.53\pm 0.01
        \end{array}$}
    && {$
        \begin{array}{c}
          0.59\pm 0.01 \\
          0.45\pm 0.01
        \end{array}$}
    && {$
        \begin{array}{c}
        2/3 \\
        $---$
        \end{array}$} \\
  {$C$}
  & {$\alpha^\prime/
        \begin{array}{c}
          \downarrow \\
          \textcolor{red}{\bullet}
        \end{array}$}
    && {$0.69 \pm 0.01$} && {$0.69 \pm 0.01$} && {$2/3$} \\
    & {$\alpha\,/
        \begin{array}{c}
        \textcolor{blue}{\bullet} \\
        \uparrow
        \end{array}$}
    && {$0.50 \pm 0.01$} && {$0.40 \pm 0.02$} && {$1/2$}  \\
\hline
\end{tabular}
\begin{flushleft}
TABLE I: Critical exponents (C.E.): the arrow $\rightarrow\textcolor{red}{\bullet}$  $\left(
\begin{array}{c}
  \textcolor{blue}{\bullet} \\
  \uparrow
\end{array}
\right)$
indicates the path in the $\mu\,(T)-$ direction to the CEP (TCP) for ${\mu<\mu^{CEP}}$ $({T<T^{TCP}}$).
\end{flushleft}
\end{center}
\end{table}
The phenomenological relevance of fluctuations in the finite temperature and chemical
potential around the CEP/TCP of QCD has been recognized by several authors.
If the critical region of the CEP is small, it is expected that most of the fluctuations
associated with the CEP will come from the mean-field region around the
CEP \cite{Hatta:2003PRD}.

\begin{figure*}[t]
    \begin{center}
    \begin{tabular}{cc}
        \hspace*{-0.5cm}\epsfig{file=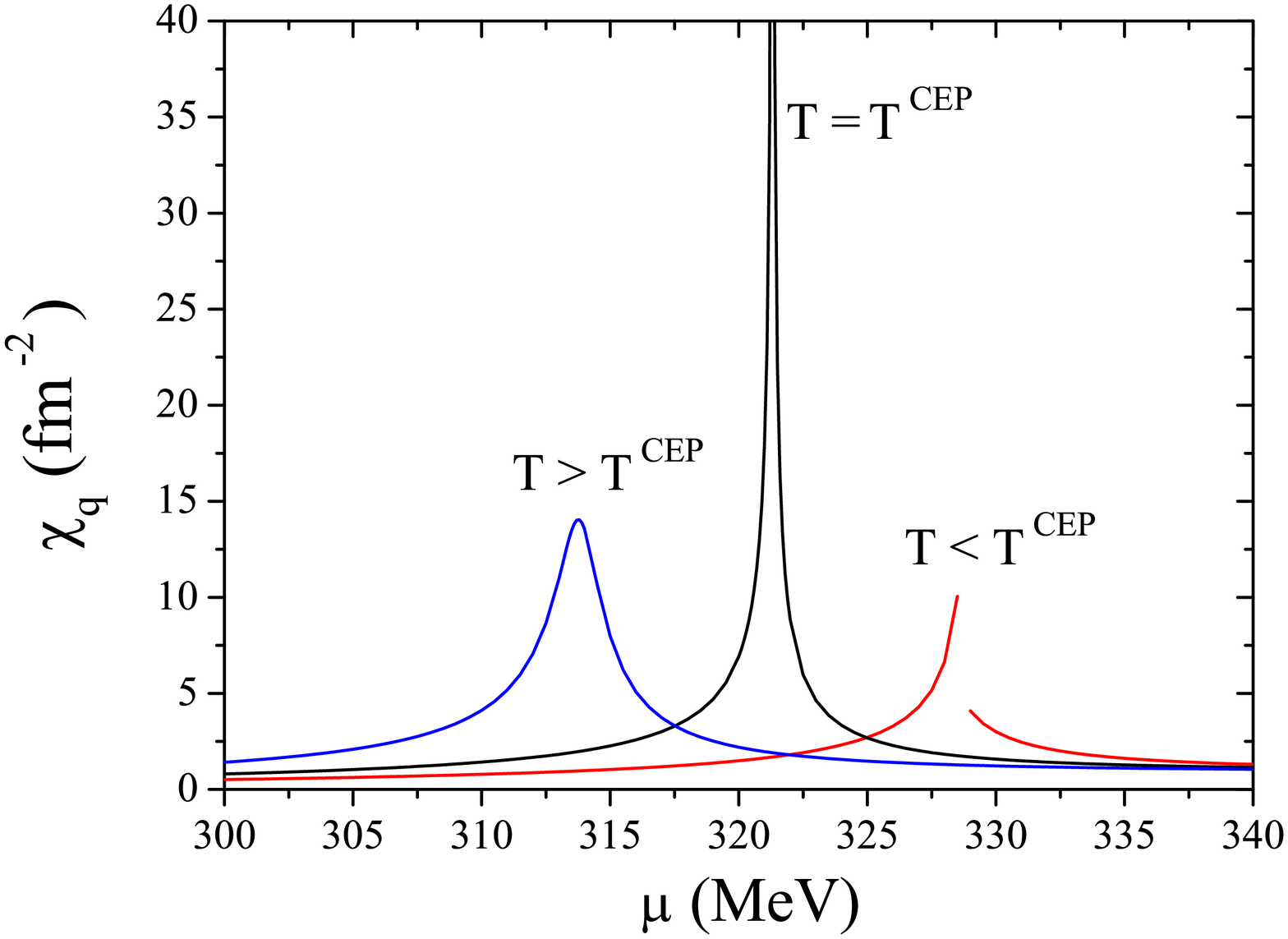,width=8.5cm,height=7.5cm} &
        \hspace*{-0.5cm}\epsfig{file=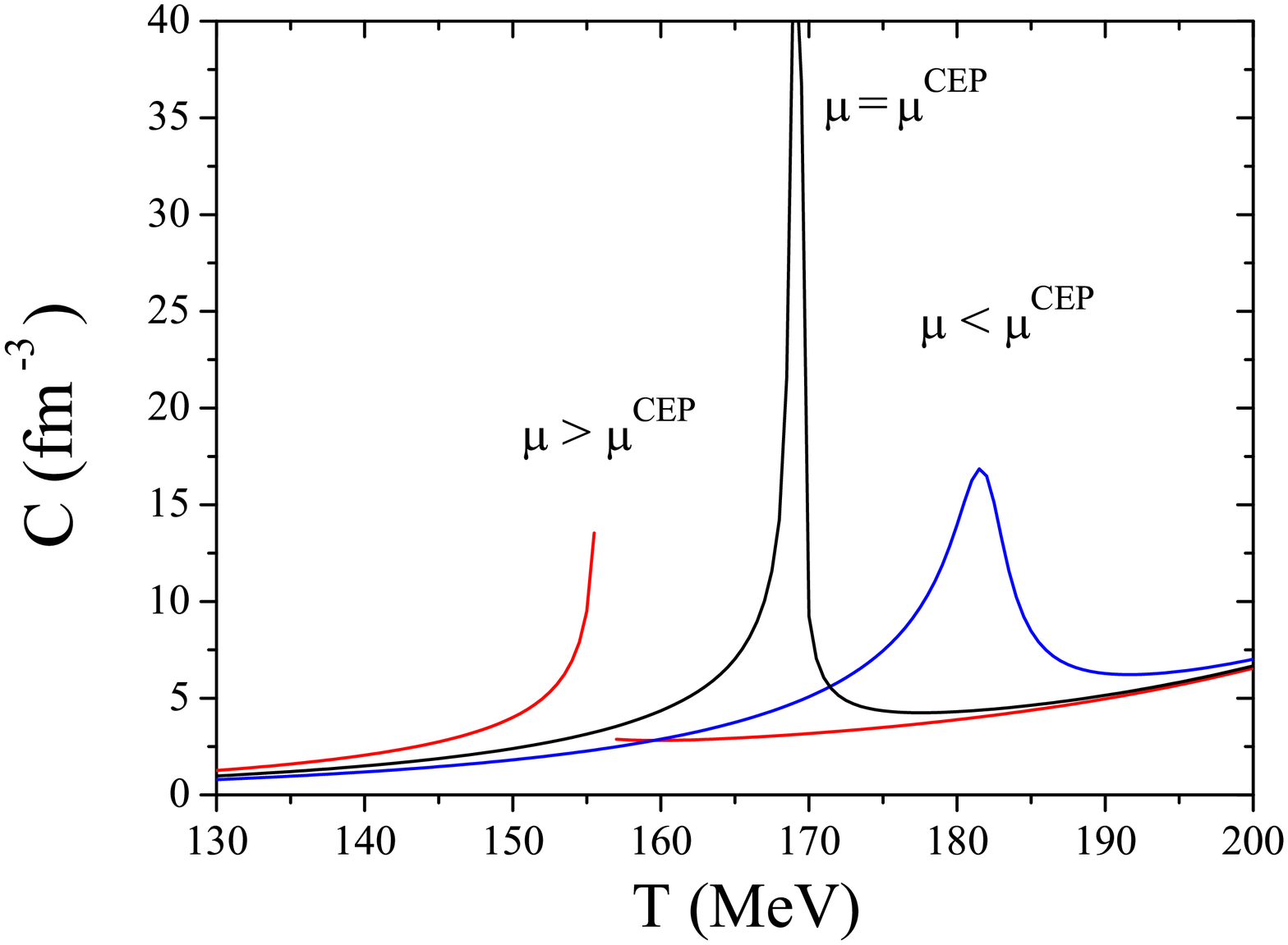,width=8.5cm,height=7.5cm} \\
    \end{tabular}
    \end{center}
\vspace{-0.5cm} \caption{Left panel: baryon number susceptibility as a function of $\mu$
for different temperatures around the CEP in PNJL model: $T^{CEP}=169.11$ MeV and $T=T^{CEP}\pm10$ MeV.
Right panel: specific heat as a function of $T$ for different values of $\mu$ around
the CEP: $\mu^{CEP}=321.32$ MeV and $\mu=\mu^{CEP}\pm10$ MeV.}
\label{Fig:2}
\end{figure*}

In the left panel of Fig. \ref{Fig:2}, $\chi_q$ is plotted
as a function of $\mu$ for three different temperatures around the CEP.
For temperatures below $T^{CEP}$ we have a first order phase transition and,
consequently, $\chi_q$ has a discontinuity. For $T = T^{CEP}$ the slope of the baryon
number density tends to infinity at $\mu=\mu^{CEP}$, which implies a diverging
$\chi_q$. For temperatures above $T^{CEP}$, in the crossover region, the
discontinuity of $\chi_q$ disappears at the transition line.

A similar behavior is found for the specific heat as a function of temperature for three
different chemical potentials around the CEP, as we can see from the right panel of
Fig. \ref{Fig:2}.

These behaviors of $\chi_q$ and $C$ are qualitatively similar to those obtained in the $SU$(2)
NJL model \cite{Costa:2007PLB}.
As we have already seen, the baryon number susceptibility, $\chi_q$, and the specific
heat, $C$, diverge at $T = T^{CEP}$ and $\mu = \mu^{CEP}$, respectively
\cite{Hatta:2003PRD,Costa:2007PLB}. In order to make this statement more
precise, we will focus on the values of the critical
exponents, in our case $\epsilon$ and $\alpha$ are the critical exponents of $\chi_q$ and
$C$, respectively.
These critical exponents will be determined by finding two directions,
temperature-like and magnetic-field--like, in the $(T-\mu)$-plane near the CEP, because,
as pointed out in \cite{Griffiths:1970PR}, the form of the divergence depends on the
route which is chosen to approach the critical end point.

\begin{figure*}[t]
\begin{center}
  \begin{tabular}{cc}
    \hspace*{-0.5cm}\epsfig{file=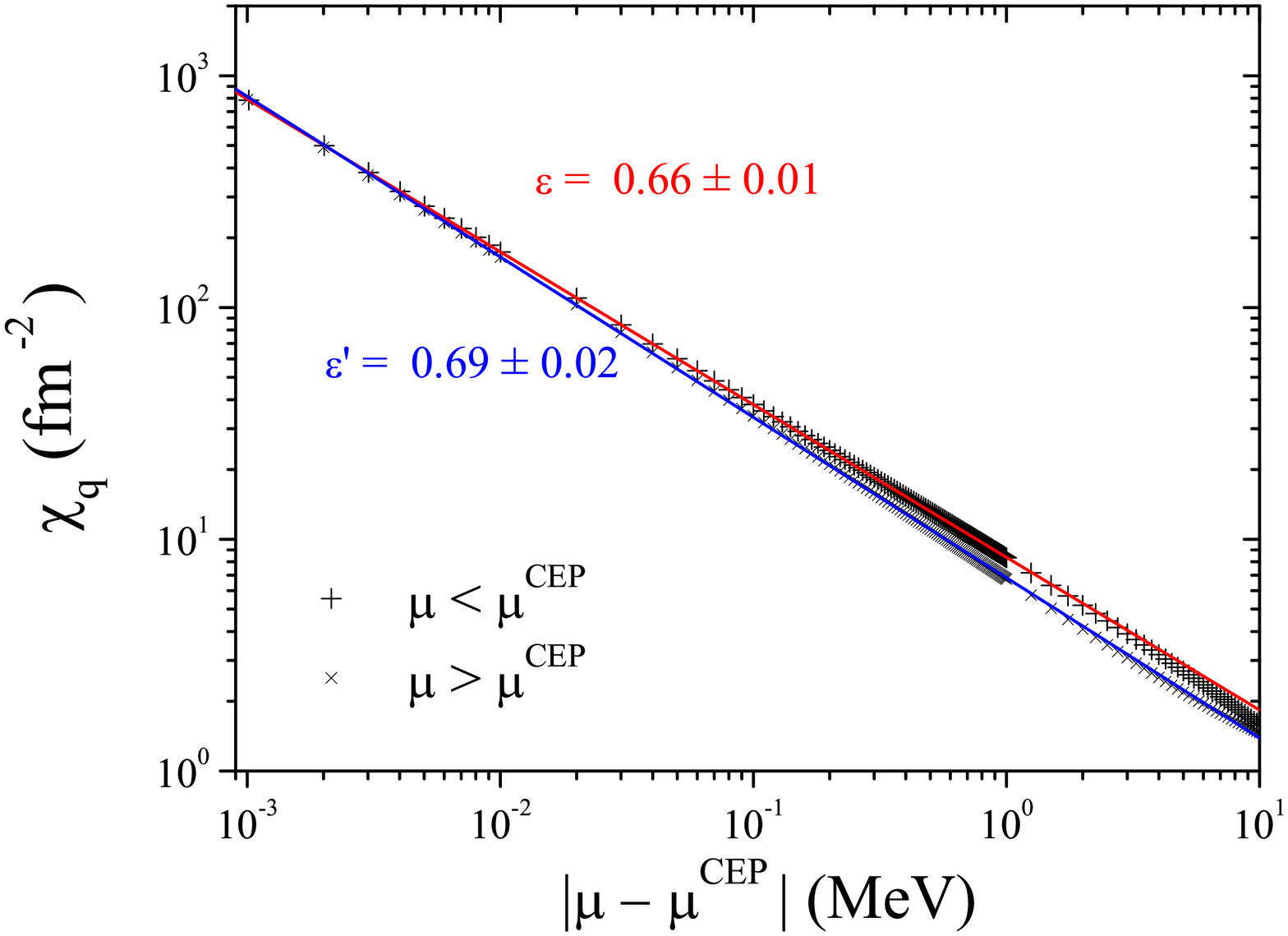,width=8.5cm,height=7.5cm} &
    \hspace*{-0.5cm}\epsfig{file=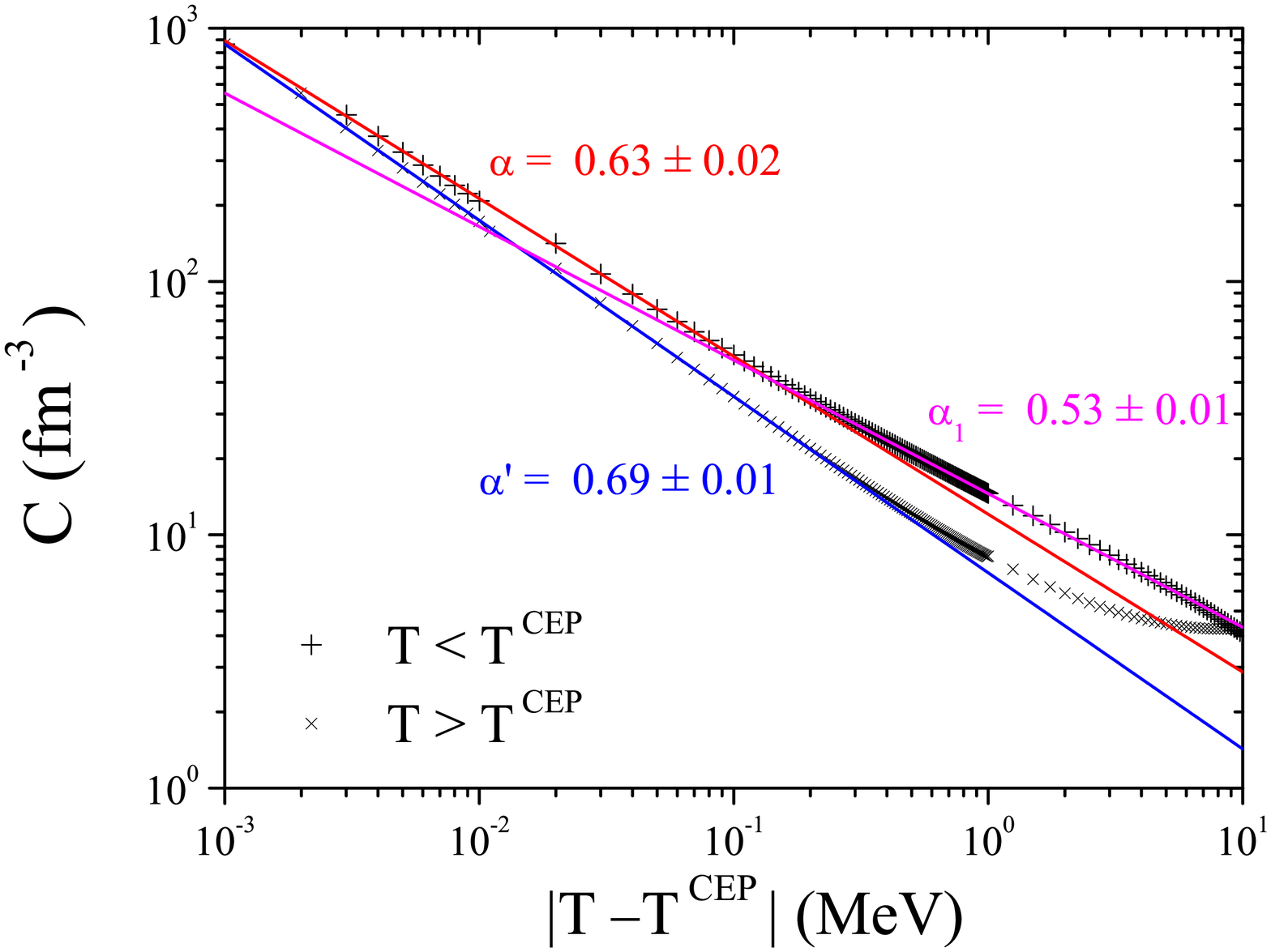,width=8.5cm,height=7.5cm} \\
  \end{tabular}
\end{center}
\vspace{-0.5cm} \caption{Left panel: baryon number susceptibility as a function of
$|\mu-\mu^{CEP}|$ at the fixed $T=T^{CEP}$. Right panel: specific heat as a function of
$|T-T^{CEP}|$ at the fixed $\mu=\mu^{CEP}$. } \label{Fig:3}
\end{figure*}

To study the critical exponents for the baryon number susceptibility (Eq. \ref{chi})
we will start with a path parallel to the $\mu$-axis in the ($T,\mu$)-plane,
from lower $\mu$ towards the critical $\mu^{CEP}=321.32$ MeV, at fixed temperature
$T^{CEP}=169.11$ MeV. In Fig. \ref{Fig:3} we plot $\chi_q$ as a function of
$\mu$ close to the CEP. Using a linear logarithmic fit
\begin{equation}
    \ln \chi_q = -\epsilon \ln |\mu -\mu^{CEP} | + c_1 ,
\end{equation}
where the term $c_1$ is independent of $\mu$, we obtain $\epsilon = 0.66\pm 0.01$,
which is consistent with the mean-field theory prediction $\epsilon = 2/3$.

We also study the baryon number susceptibility from higher $\mu$ towards the critical
$\mu^{CEP}$. The logarithmic fit used now is $\ln \chi_q = -\epsilon' \ln |\mu
-\mu^{CEP}| + c'_1$. Our result shows that $\epsilon'  = 0.69\pm 0.02\approx \epsilon$. This
means that the size of the region we observe is approximately the same independently of
the direction we choose for the path parallel to the $\mu$-axis. These critical exponents
are presented in Table I, where  we can see that the critical exponents for the
baryon number susceptibility are approximately the same for both, PNJL and NJL models,
and are consistent with the mean-field theory prediction $\epsilon = 2/3$.

On the other hand, in the chiral limit (where the CEP becomes a TCP), it is  found that
the critical exponent for $\chi_q$ has the value $\gamma_q=0.51\pm0.01$, for
both, PNJL and NJL models. Again, these results are in agreement with the mean-field value
($\gamma_q=1/2$).

Now, paying attention to the specific heat around the CEP, we have used a path parallel
to the $T$-axis in the ($T,\mu$)-plane from lower (higher) $T$ towards the critical
$T^{CEP}=169.11$ MeV at fixed $\mu^{CEP}=321.32$ MeV. In Fig. \ref{Fig:3} (right panel)
we plot $C$ as a function of $T$ close to the CEP in a logarithmic scale.
We see that for the region $T<T^{CEP}$ we have $\alpha=0.63\pm 0.02$\footnote{We use
the linear logarithmic fit $\ln C = -\alpha \ln |T -T^{CEP}| + c_2$ where the term $c_2$
is independent of $T$.}. Contrarily to what happens in the NJL model (see Table I and
Refs. \cite{Costa:2007PLB}), this value of $\alpha$ is closer to the one
suggested by universality arguments in \cite{Hatta:2003PRD}.

We also observe, as in in Ref. \cite{Costa:2007PLB} that, in PNJL (NJL) model, for the
region $T<T^{CEP}$ we have a slope of data points that changes for values of
$|T-T^{CEP}|$ around $0.3$ MeV. We have fitted the data for $|T-T^{CEP}|<0.3$ MeV and
$|T-T^{CEP}|>0.3$ MeV separately and obtained, respectively, the critical exponent
$\alpha=0.63\pm 0.02$ ($\alpha=0.59\pm 0.01$) and $\alpha_1=0.53\pm 0.01$
($\alpha_1=0.45\pm 0.01$), which have a linear behavior for several orders of magnitude
(see Table I). As pointed out in \cite{Hatta:2003PRD}, this change of the
exponent can be interpreted as a crossover of
different universality classes, with the CEP being affected by the TCP.

In both models, the influence of the TCP is stronger in the specific heat rather than in
the baryon number susceptibility: the closest distances between the TCP and the CEP in
both phase diagrams occur in the T-direction ($(T^{TCP} - T^{CEP})<(\mu^{CEP} - \mu^{TCP})$).
When the CEP is approached from above the trivial exponent
$\alpha^\prime =0.69$ (for both models) is obtained.

Let us now analyze the behavior of the specific heat around the TCP.
As shown in Table I, we find a nontrivial critical exponent
$\alpha=0.40\pm0.02$ only for the NJL model while for the PNJL model
$\alpha=0.50\pm0.01$.


In this work, we have considered an extension of the NJL model which couples chiral
and confinement-like order parameters.
We have found that our model in general reproduces important features of the QCD
phase diagram as the location of the CEP/TCP.
In addition, these results confirm the general idea that, in contrast to the NJL model, 
the PNJL model provides a
quantitative description of QCD thermodynamics near critical points.
In the PNJL model, the crossover taking place in a smaller $T$ range can be interpreted
as a crossover transition closer to a second order one than in the NJL
model.
This ``faster'' crossover may explain the elongation of the critical region
compared to the NJL one
giving rise to a greater correlation length even far from
the CEP.
We have also studied the baryon number susceptibility and the specific heat around the
CEP which are related with event-by-event fluctuations of $\mu$ or $T$ in heavy-ion
collisions. An important observation is that, in the PNJL model the obtained critical
exponents are consistent with the mean-field values, both for the baryon number
susceptibility and the specific heat, while for the NJL this is only true for the baryonic
susceptibility, since  for the specific heat $\alpha$ is different from $\epsilon$.

As the CEP lies in the region expected to be probed by heavy ion experiments, it
would be interesting to find an experimental signature of such a point. Near the critical
point, and in particular in the path we choose to study the critical exponents of the
specific heat, there is a possibility of the spinodal decomposition in the first order
phase transition. So, the competition between features of the first and second order
phase transition in the mixed phase can allow for nontrivial effects, such as the above
referred ones and to which there is no information from heavy-ion collisions. Our numerical
results which also includes the chemical potential can be relevant to this purpose.

In conclusion, the results with the PNJL model are closer to lattice results and we also
recover the universal behavior of the critical exponents of both the baryon
susceptibility and the specific heat.  The PNJL model here discussed, allowing for finite
dynamical quark masses, can provide a convenient tool to study the QCD phase diagram; it
allows to establish a convenient link between the lattice results, and the NJL model
itself where gluonic degrees of freedom are missing.

\begin{acknowledgments}
Work supported by grant SFRH/BPD/23252/2005 (P. Costa), Centro de
F\'{\i}sica Te\'orica, FCT under projects POCI/FP/63945/2005 and
POCI/FP/81936/2007 (H. Hansen).
This work was done in spite of the lack of support from Ministry of University
and Research of France.

\end{acknowledgments}


\end{document}